\documentclass[useAMS,usenatbib]{mn2e}

\usepackage{graphicx,psfig}

\usepackage{psfig}

\def\ergscm2{erg\,s$^{-1}$cm$^{-2}$}

\def\uu{4U\,0142+614\,}
\def\ee{1E\,1048-5937\,}
\def\kes{1E\,1841-045\,}

\def\rxs{1RXS\,J1708-4009\,}
\def\ea{1E\,2259+586\,}
\def\xte{XTE\,J1810-197\,}
\def\cxo{CXOU\,J0100-7211\,}
\def\wes{CXOU\,J1647-4552\,}
\def\1e{1E\,1547.0-5408\,}
\def\sgra{SGR\,1806-20\,}
\def\sgrb{SGR\,1900+14\,}

\def\sgrd{SGR\,1627-41\,}

\newcommand{\XMM}{{\it XMM--Newton}\,}

\newcommand{\INT}{{\it INTEGRAL}\,}
\newcommand{\CXO}{{\it Chandra}\,}

\title[X-ray spectra from magnetar candidates - III.]{
X-ray spectra from magnetar candidates - III.
Fitting SGRs/AXPs soft X-ray emission with
non-relativistic Monte Carlo models}
\author[ ]{S. Zane$^{1}$
\thanks{E-mail:
sz@mssl.ucl.ac.uk (SZ);
n.rea@uva.nl (NR); turolla@pd.infn.it (RT);
nobili@pd.infn.it (LN).},
N. Rea$^{2}$, R.
Turolla$^{3, 1}$ and L. Nobili$^{3}$\\
$^{1}$ Mullard Space Science Laboratory,
University College London, Holmbury St. Mary, Dorking, Surrey, RH5 6NT,
UK; \\
$^2$ University of Amsterdam, Astronomical Institute ``Anton
Pannekoek'',
Kruislaan, 403, 1098~SJ, Amsterdam, The Netherlands; \\
$^3$ Department of Physics, University of Padova, via
Marzolo 8, 35131 Padova, Italy.}

\begin{document}

\date{}

%\pagerange{\pageref{firstpage}--\pageref{lastpage}} \pubyear{2007}

\maketitle

\label{firstpage}

\begin{abstract} Within the magnetar scenario, the ``twisted 
magnetosphere'' model appears very promising in explaining the persistent 
X-ray emission from the Soft Gamma Repeaters and the Anomalous X-ray 
Pulsars (SGRs and AXPs). In the first two papers of the series, we have 
presented a 3D Monte Carlo code for solving radiation transport as soft, 
thermal photons emitted by the star surface are resonantly upscattered by 
the magnetospheric particles. A spectral model archive has been generated 
and implemented in {\tt XSPEC}. Here we report on the systematic 
application of our spectral model to different \XMM\ and \INT\ 
observations of SGRs and AXPs. We find that the synthetic spectra provide 
a very good fit to the data for the nearly all the source (and source 
states) we have analyzed.

\end{abstract}


\section{Introduction}
\label{intro}

Over the last few years, increasing observational evidence has
gathered in favour of the existence of ``magnetars'', i.e. neutron
stars (NSs) endowed with an ultra-strong magnetic field ($B \approx
10^{14}-10^{15}$~G), about ten times higher than 
the critical threshold at which
quantum electro-dynamical (QED) effects become important ($B_{crit}
\sim 4.4 \times 10^{13}$~G). The family of magnetars candidates
comprises two classes of sources, the anomalous X-ray pulsars (AXPs)
and the soft $\gamma$-ray repeaters (SGRs).  About fifteen objects are
known, all characterized by similar properties: slow X-ray pulsations
($P\sim2$--12\,s), large spin-down rates ($\dot P\sim
10^{-13}$--$10^{-10}$ s/s), a typical persistent X-ray luminosity of
$\approx 10^{34}$--10$^{36}\ {\rm erg\, s}^{-1}$, 
lack of bright optical companions (favouring an interpretation in terms of 
isolated objects), and a high
level of bursting/flaring activity which can differ among the two
classes \cite[see][for recent reviews.]{woodsrew,mere08}

Spectral data can provide key information about the physics of 
ultra-magnetized NSs. High-energy observations of SGRs and AXPs persistent
emission cover now the $\sim 0.5$--10 (\XMM/\CXO) and $\sim 20$--200 
(\INT) keV bands. It is  possible, although not proved yet, that
different emission mechanisms are responsible 
for the emission in the two bands. 

The low-energy ($\la 10$ keV) spectrum of the persistent 
(i.e. outside bursts) emission is often modelled in terms of
a blackbody ($kT\sim 0.3$--0.6 keV) plus a power-law with photon 
index $\Gamma_s\sim 2$--4,
although in some AXPs a two blackbody model has been also applied. 
The X-ray persistent emission
above 20 keV has a power-law spectral shape ($\Gamma_h\sim 2$) which, in 
particular in AXPs, is markedly harder than that observed below 10 keV 
\cite[see again][and references therein]{mere08}. However, although these 
phenomenological fitting models have been systematically applied over 
the last decade, a convincing physical interpretation of the various spectral 
components is still missing.

In the magnetar framework, the twisted magnetosphere 
model \citep{dt92, td93, tlk02} offers a promising physical 
interpretation for the X-ray emission from SGRs/AXPs. In particular, the 
supporting currents required to sustain the twist are substantially in 
excess of the Goldreich-Julian current and can produce large optical depth 
to resonant cyclotron scattering (RCS). Soft
(thermal) photons produced at the star surface gain energy in repeated 
scatterings and this leads to the formation of an extended, high-energy 
tail, superimposed to the seed thermal component. The qualitative predictions 
of this model have been verified to match some spectral and timing properties  
of magnetar sources, as the spectral shape observed in quiescence below 
$\sim10$~keV and the long term variation observed in some sources 
\cite[e.g.][]{sandro05,rea05,cam07}. 

More recently, several efforts have been carried out recently in order
to test the model  quantitatively against real data using different
approaches and with a varying degree of sophistication. 
A first attempt in this direction has been presented by
\cite{lg06}.  These authors developed a very simple, semi-analytical
treatment of the RCS process by working in 1-dimensional geometry.
They assumed that seed photons are
emitted by the NS surface with a blackbody spectrum, and Thomson
scattering occurs in a thin, plane parallel magnetospheric slab
permeated by a static, non-relativistic, warm medium at constant
electron density. 
These models have been implemented in
XSPEC by \cite{nanda1,nanda3,nanda2} and successfully applied to all
magnetars spectra below 10keV. The good agreement
found, even within a simplified treatment, supports the idea that RCS in 
a sheared magnetosphere
plays a central role is the formation of magnetar spectra in the (soft) X-ray
range. The same RCS
model has been used by \cite{guv07}, who assumed that seed photons comes from
an atmosphere surrounding the star. More 
recently, 3-D Monte Carlo calculations have been presented by
\cite{ft07}, although these spectra have never been applied to fit
X-ray observations.

Motivated by this, we present here a systematic application of our 
3D Monte Carlo
spectral calculations \cite[see][for all details]{ntz1, ntz2} to AXPs and 
SGRs spectral data. The paper is organized as follows. In \S~\ref{mod} we 
briefly
summarize the basic features of the model.  The data sample
and spectral results are presented in \S~\ref{appl}. Discussion and
conclusions follow in \S\S~\ref{disc}, \ref{conc}.

\section{The model} \label{mod}

As discussed in detail in \cite{ntz1,ntz2}, we have
recently developed a 3-dimensional treatment of RCS, aimed at a
detailed investigation of the spectral, timing and polarization
properties of magnetars \cite[see also][]{pav08}. 
Our Monte
Carlo code, which  is completely general and can handle different
magnetic field topologies and distributions of seed
photons, has been used to produce an archive of spectral models 
that have been subsequently implemented in XSPEC. 
Such models
rely on a number of choices for the assumed configuration, that are 
discussed in detail in
\cite{ntz1} and briefly summarized in this section.

First, the XSPEC archive  models have been computed by assuming that
the star surface emits as a blackbody at an uniform temperature, 
$kT$, and that the surface radiation
is isotropic and unpolarized. The magnetic field topology is assumed
to be a twisted, force free dipole and is uniquely
characterized by the value of the twist angle, $\Delta \phi$
\cite[see][]{tlk02}. No attempt is made to fit the value of the
polar field strength, that has been fixed at $10^{14}$~G. Our model
is based on a simplified treatment of the charge carriers velocity
distribution which accounts for the particle collective motion, in
addition to the thermal one. Pair production has been
neglected. Magnetospheric electrons stream freely along
the field lines  (the motion across the field lines is quantized).
The electron velocity distribution parallel to the field is taken to be a 1-D
relativistic Maxwellian at temperature $T_e$,
superimposed to a bulk motion with velocity $\beta_{bulk}$ (in units
of the light velocity $c$). Besides, in order to minimize the number
of free parameters,  the models in the archive were computed 
assuming that  the electron temperature is related to  $\beta_{bulk}$ 
\cite[see][for all details]{ntz1}.  
Scattering in a magnetized medium was treated by considering
only the resonant part of the magnetic Thomson cross section
and neglecting electron recoil along the field direction. For 
the sake of conciseness, in the following this approximation 
will be referred to as the (resonant) Thomson limit. On the other hand, 
the code is 
completely general and inclusion of the relativistic 
QED resonant cross section, which is required in the modelling of the
hard ($\sim 20$--200 keV) spectral tails observed in the magnetar
candidates, is under way \cite[see][]{ntz2,pav08}. 
Electron recoil in the direction parallel to the field 
starts to be important when the photon energy in the electron rest
frame becomes comparable to the electron rest energy. If $\gamma$ is
the mean electron Lorentz factor, this occurs at typical energies
$\sim m_ec^2/\gamma$. Assuming mildly relativistic particles, the
previous limit implies that conservative scattering should provide
good accuracy up to photon energies of some tens of keV. This,
together with the fact that resonant scattering occurs in regions
where $B\ll B_{crit}$, makes the use of the (much simpler)
non-relativistic (Thomson) magnetic cross section adequate. 
The final XSPEC {\tt atable} spectral model (22~MB in size, 
named {\tt ntznoang.mod},
hereafter NTZ) has been created by using the routine {\tt wftbmd},
available on-line.\footnote{see
  http://heasarc.gsfc.nasa.gov/docs/heasarc/ofwg/docs/general/ \\
  modelfiles\_memo/modelfiles\_memo.html.} In summary, it depends on
four free parameters ($\beta_{bulk}$, $\Delta \phi$, $\log kT$ plus a
normalization constant), which can be simultaneously varied during the
spectral fitting following the standard $\chi^2$ minimization
technique. In NTZ models
the photon number is conserved and the monochromatic number flux
which reaches infinity is the same as that of the seed blackbody
spectrum\footnote{Photon number is not necessary conserved when
Landau-Raman scattering is accounted for (photon spawning;
\citealt{ntz2}). Present models have been obtained under the
assumption of conservative magnetic scattering, so photon number
conservation is ensured. Also, spectra in the model archive have
been computed averaging over all viewing directions, i.e. photons
are collected over the entire observer's sky. When one particular
line-of-sight is chosen, the photons reaching the observer are
clearly less than those emitted by the surface.}. 
This implies
that the normalization constant (norm) divided by $kT^3$ is
proportional to the emitting area on the star surface (see \S\ref{corr}).  
It is important to note that the NTZ model has the same number
of free parameters than the canonical blackbody plus power-law
empirical model or the 1D RCS model recently discussed in
\cite{nanda2}, and hence has the same statistical significance. In the
following sections we present its systematic application to the soft
(0.5--10 keV) X-ray spectra of magnetar candidates.

\section{Application to magnetar's soft X-ray spectra}
\label{appl}

We applied the NTZ model to a large sample of AXPs and SGRs, using
\XMM\, and \INT\, data. The sample of datasets basically coincides
with that used in \cite{nanda2}, and we refer to that paper for more
details and for a discussion on the data analysis. However, at
variance with \cite{nanda2}, we did not consider here the transient
magnetars \xte\, and \wes, the analysis of which will be
reported in a separate paper devoted to a detailed investigation of
their outburst evolution over a period of several years. 
On the other hand, we included the recent \XMM\, observations of
\cxo\, and \sgrd, that were not available at the time of the previous
investigation \cite[see][for further details on these
observations]{tiengo08,paolo08}.

All fits have been performed using {\tt XSPEC} version 11.3 and 12.0, for 
a consistency check. A 2\% systematic error was added to the data to 
partially account for uncertainties in instrumental calibrations. Only in 
the case of \cxo, which has a very low absorbtion, the fit has been 
performed in the 0.5--10\,keV energy range.  For the other sources, that 
are highly absorbed, the 0.5--1\,keV energy range (0.5--2~keV energy range 
in the case of \sgrd) was excluded from our spectral fitting.\footnote{ 
The emission in this energy range is in fact mostly affected by 
interstellar absorption. Moreover, this is the band where most of the 
calibration issues lay \citep{hab04}.} We then checked that, for all our 
targets, the values of $N_H$ derived fitting the 1--10\,keV EPIC-pn 
spectra, are consistent (within $1\,\sigma$) with those obtained using the 
0.5--10\,keV range relative to the same data set. We used the more updated 
solar abundances by \cite{lo03}, instead of the older ones from 
\cite{ag89}. As a consequence, the value of the absorption is, on average, 
slightly higher than that reported in the literature for the same model.  
This does not affect the other spectral parameters.

\begin{figure*}
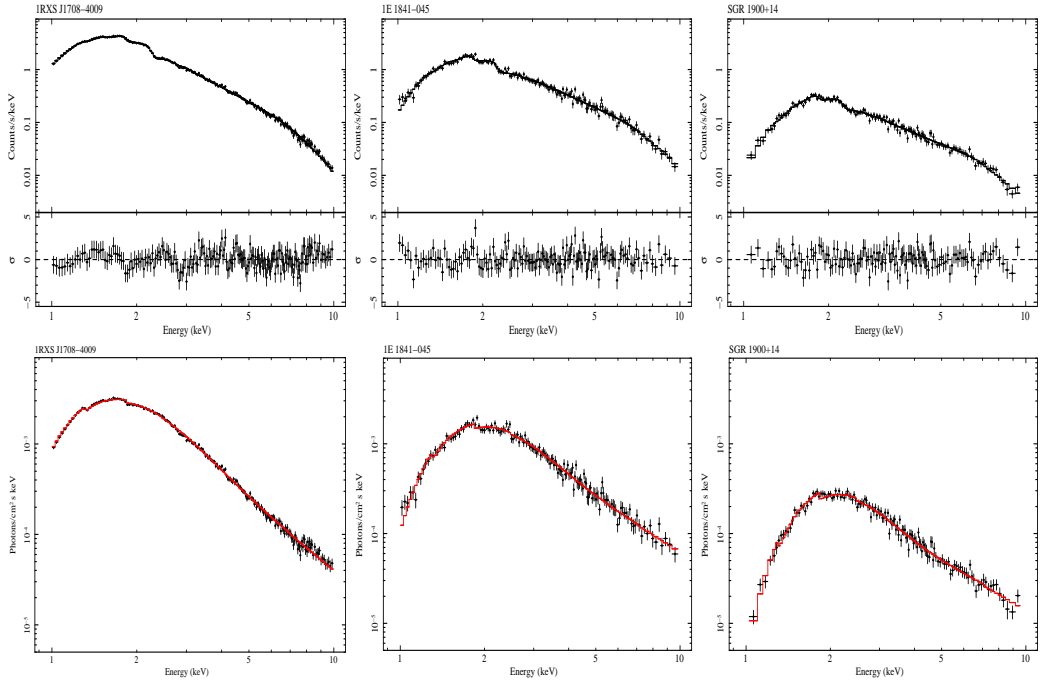

\centering{
\hspace{0.1cm}
\vbox{
\hbox{
\psfig{figure=1708_ntz_nopl_spec.ps,height=4.5cm,angle=270,width=4.5cm}
\psfig{figure=1841_ntz_nopl_spec.ps,height=4.5cm,angle=270,width=4.5cm}
\psfig{figure=1900_ntz_nopl_spec.ps,height=4.5cm,angle=270,width=4.5cm}}
\hbox{
\psfig{figure=1708_ntz_nopl_ufs.ps,height=4.5cm,angle=270,width=4.5cm}
\psfig{figure=1841_ntz_nopl_ufs.ps,height=4.5cm,angle=270,width=4.5cm}
\psfig{figure=1900_ntz_nopl_ufs.ps,height=4.5cm,angle=270,width=4.5cm}}}}
\caption{\rxs, \kes, and \sgrb: first row shows the spectra in
  Count/s/keV while in the second row we
report the photon density plots of the modelling with the NTZ model.
Only \XMM\ data have been used, and the
fitting has been restricted to the
$1-10$~keV range.
See
Table\,\ref{tablexmmnonvar} and \S\,\ref{appl} for
  details. \label{spectraxmmnonvar1}}
\end{figure*}

\begin{figure*}
\centering{
\hspace{0.1cm}
\vbox{
\hbox{
\psfig{figure=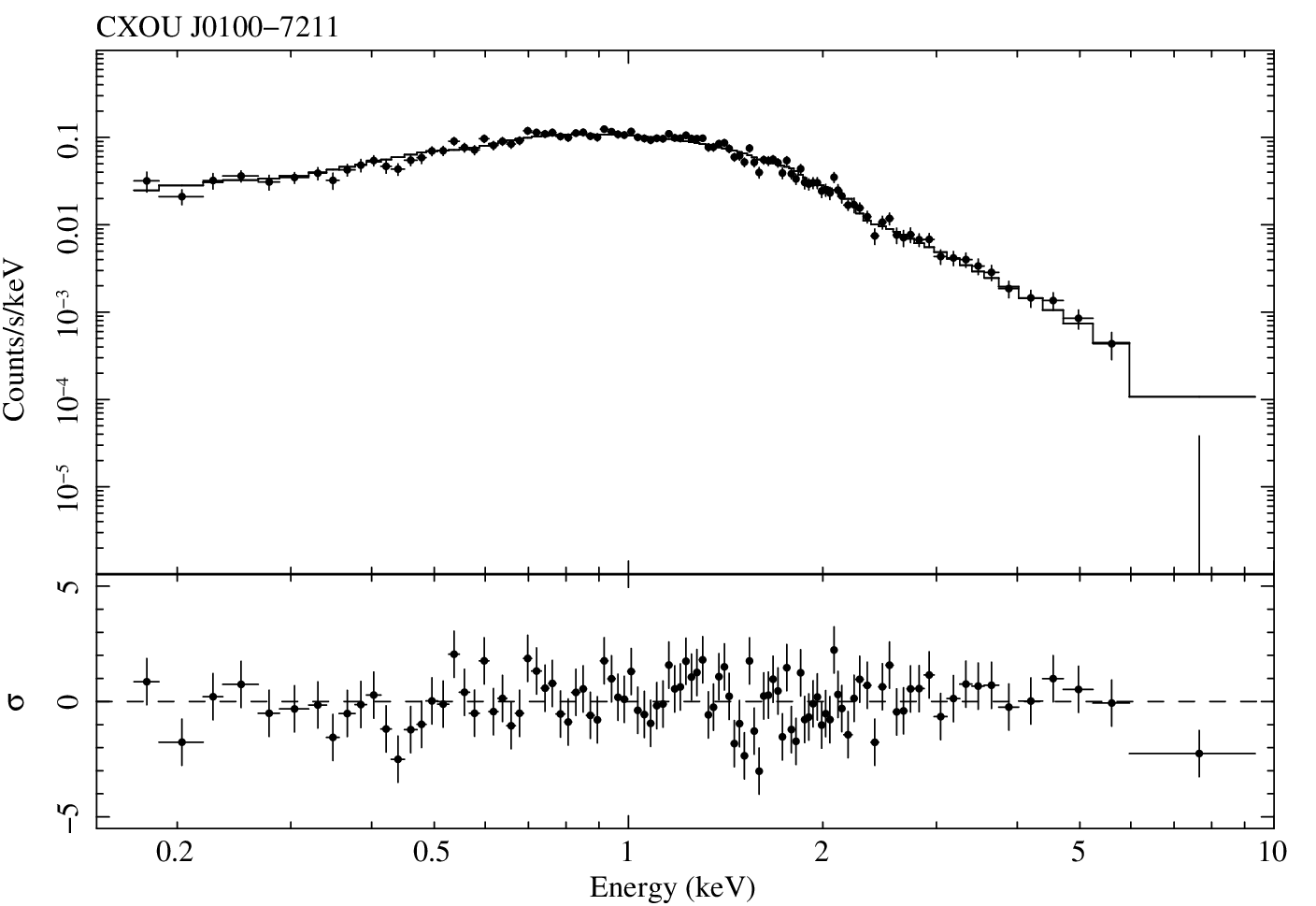,height=4.5cm,angle=270,width=4.5cm}
\psfig{figure=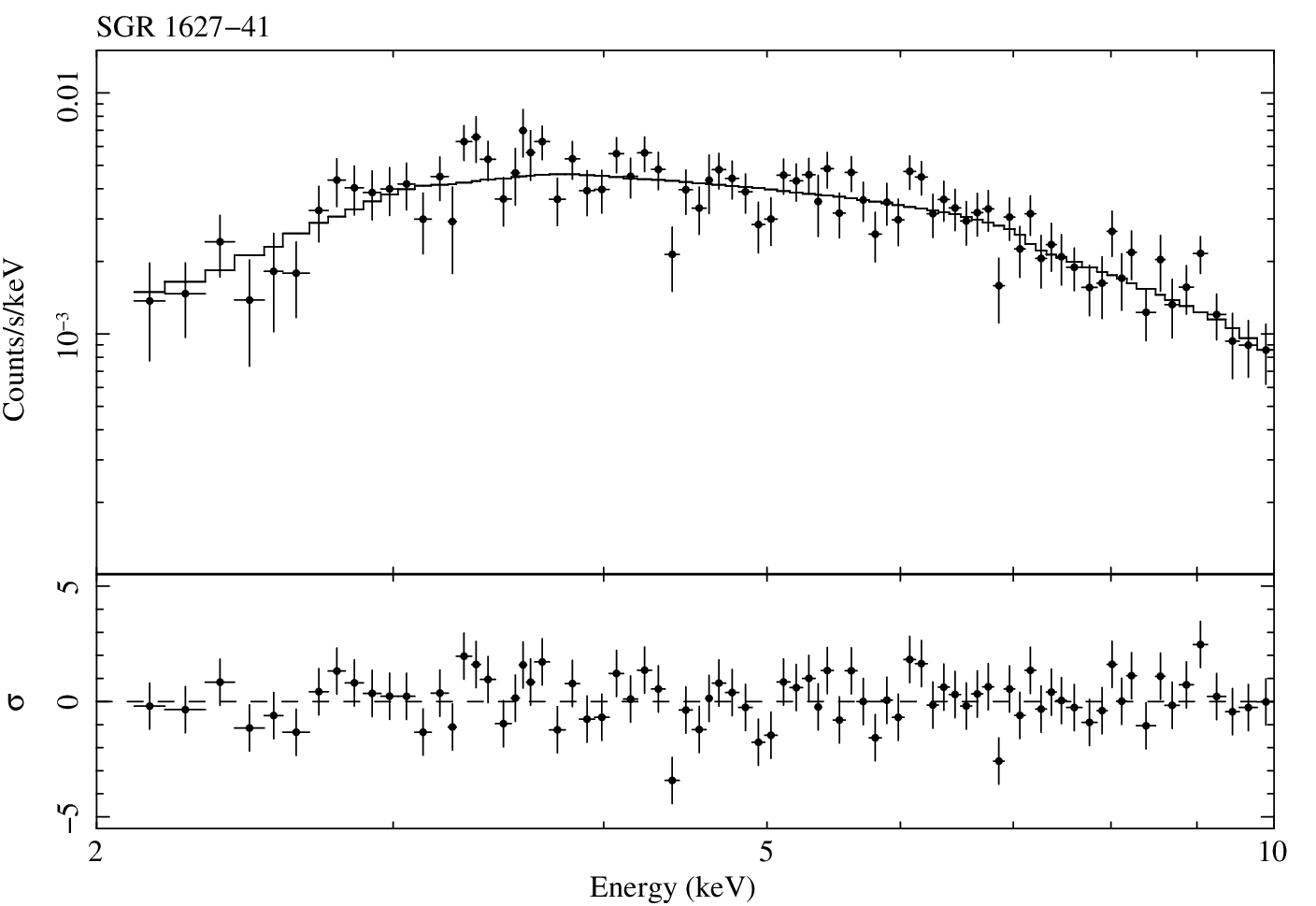,height=4.5cm,angle=270,width=4.5cm}}
\hbox{
\psfig{figure=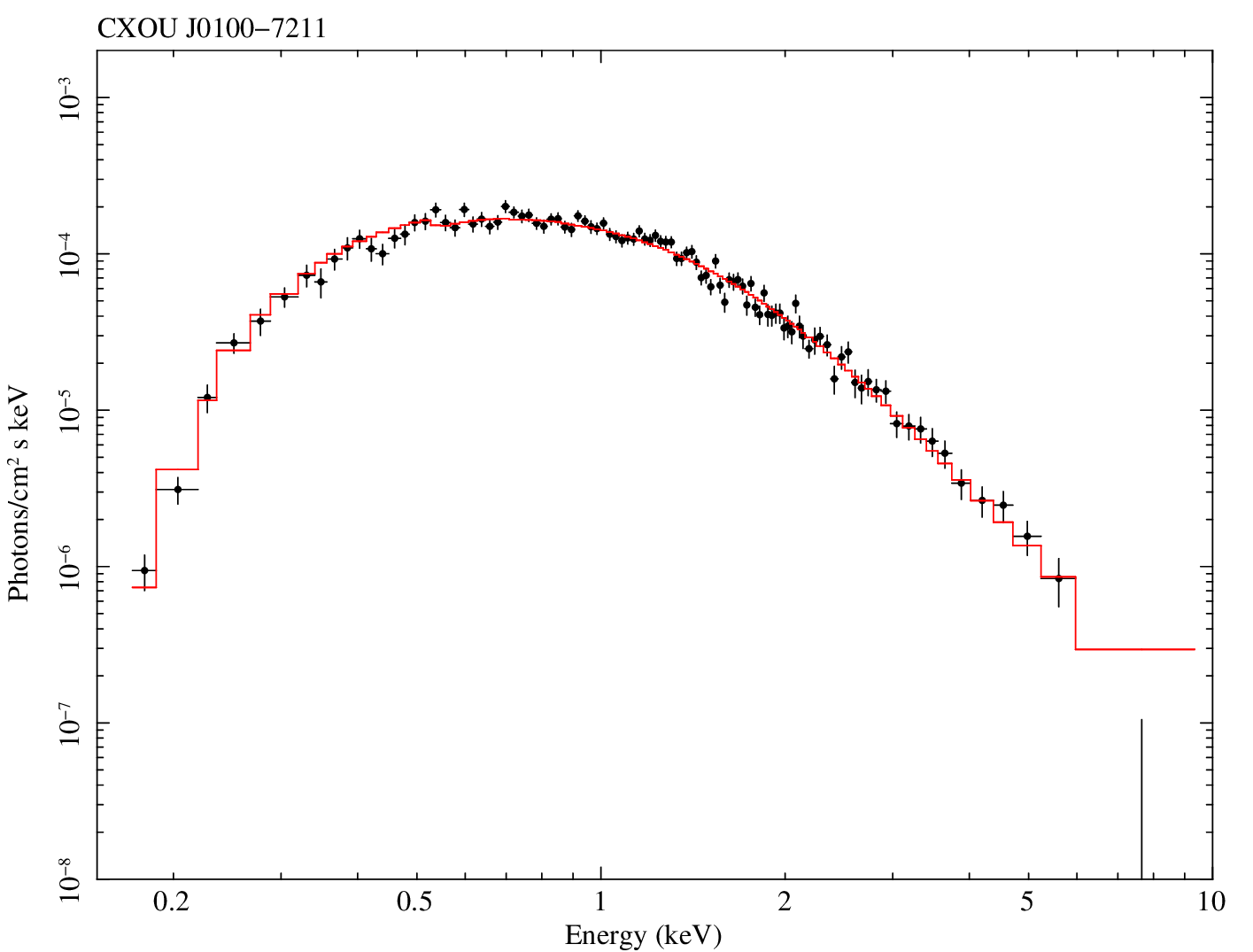,height=4.5cm,angle=270,width=4.5cm}
\psfig{figure=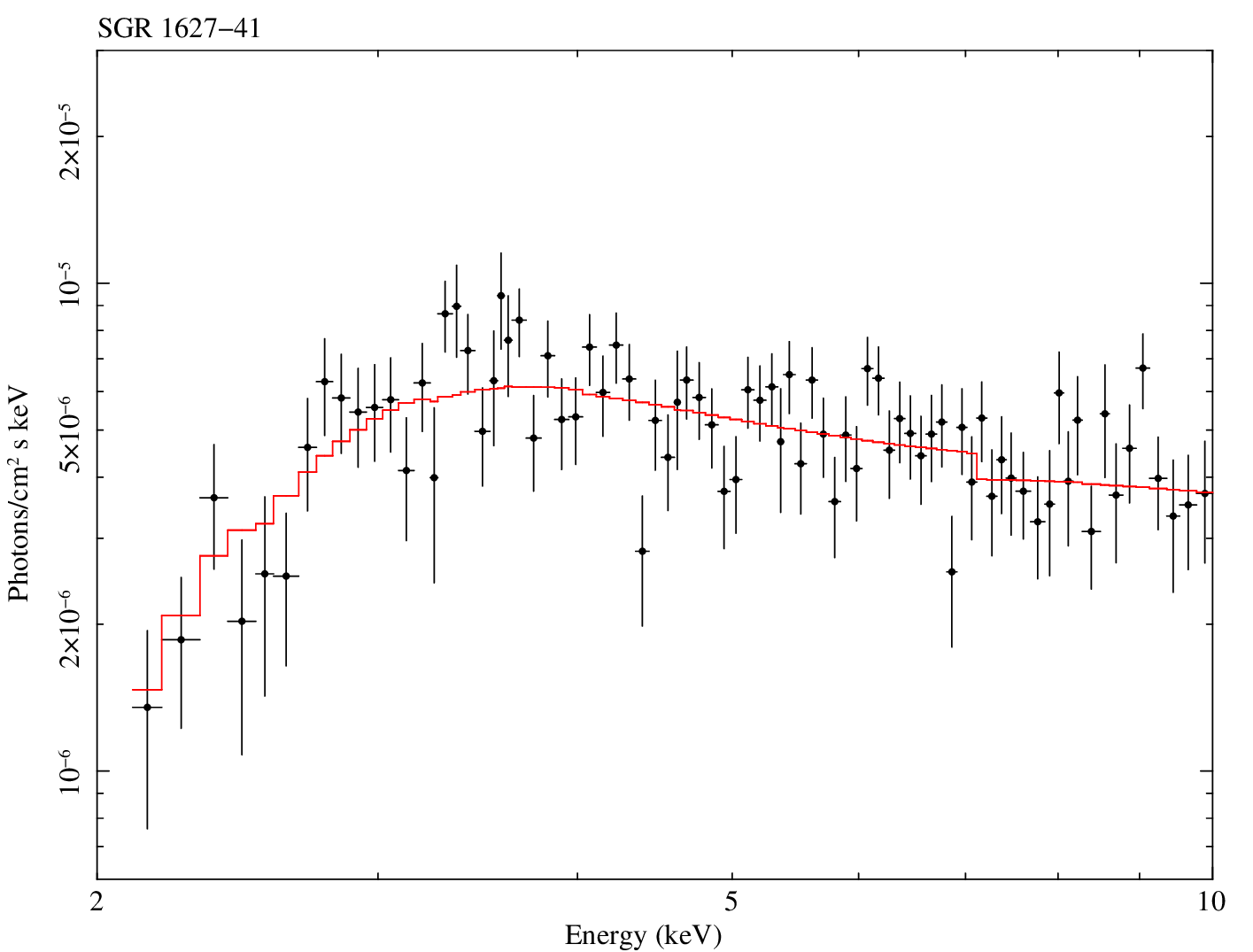,height=4.5cm,angle=270,width=4.5cm}}}}
\caption{Same as in Fig.~\ref{spectraxmmnonvar1} for 
\cxo\, and \sgrd. See
  Table~\ref{tablexmmnonvar}  for details.}
\label{spectraxmmnonvar2}
\end{figure*}

Table~\ref{tablexmmnonvar} and Figs.~\ref{spectraxmmnonvar1}, 
\ref{spectraxmmnonvar2} report our results in the 1--10~keV range for 
those sources that do not show a substantial spectral variation, in which 
cases only one dataset (the longest available) has been considered.  
Also, in Tables~\ref{tablexmmvar1}, \ref{tablexmmvar2} and 
Fig.\,\ref{spectraxmmvar} we show the fits for sources that do exhibit 
spectral variation, in which cases we considered a set of two--three 
observations for each source, corresponding to different spectral states.  
At variance with \cite{nanda2}, in these cases the different datasets have 
been fitted by imposing that the absorption, $N_H$, is the same. As it can 
be seen from the tables, in most of the cases we found that a NTZ model 
alone successfully reproduces the soft X-ray part of the spectrum up to 
10~keV, without the need of further components (see \S\ref{disc} for a 
discussion). With reference to the sample considered in \cite{nanda2}, the 
only two sources for which we do not find a satisfactory fit are \ea\ and 
\uu, which are discussed separately below.

There are a few AXPs and SGRs in our sample (namely \rxs, \kes, \uu, \ea, 
\sgrb and \sgra) that are known to have a conspicuous emission in the hard 
X-ray band. For these objects we have repeated our modelling including 
also their \INT\, spectra in order to investigate whether our model can 
reproduce part of the emission when considering the whole SED\footnote{In 
the case of \sgra\ there are no observations taken at similar epochs in 
the hard and soft X-ray bands. For this reason, and because of the high 
variability of this source, no attempt has been made to fit its broadband 
distribution either here nor in \cite{nanda2}.}. A further power-law (PL) 
has been introduced in order to account for the non thermal hard X-ray 
component, although at a purely phenomenological level.  A free constant 
was multiplied when using both \XMM\, and \INT\, data to account for 
inter-calibration uncertainties (the values of the constant was always 
differing by less than 10\% with respect to \XMM\, which was set to 
unity). Results are reported in Table~\ref{tableintegral} and 
Fig.~\ref{spectraxmmintegral}. We find that, again with the exception of 
\ea\ and \uu, in all other cases a NTZ+PL spectral decomposition 
successfully models the 1--200\,keV emission. However, the fit converges 
with a hard X-ray component that gives a substantial contribution when 
extrapolated to the soft X-ray band and, consequently, the best fitting 
parameters of the NTZ model are substantially different from those we 
found fitting the 1-10~keV emission only.

As the tables show, in some cases the fit converges with a value of 
$\Delta \phi$ which is too close to the upper bound of our model archive 
($\Delta \phi =2$), making impossible to compute the whole (closed) 
$1\sigma$ contour level. Indeed, in the case of the combined \XMM\ and 
\INT\ fits of \kes\ and \sgrb, the parameters of the NTZ model becomes 
less constrained than when fitting the \XMM\ data alone, and we could only 
set an upper limit on $\Delta \phi$ (see Table\,\ref{tableintegral}). 
Although we do not regard this as a particular problem, we should caveat 
that in some fits the value of the twist angle appears to be less 
constrained than those of the other model parameters.

\section{Discussion}
\label{disc}

In this paper we have applied a 3D Monte Carlo model of resonant 
scattering to the phase averaged spectra of an extensive set of magnetar 
sources. The sample is the same as in \cite{nanda2}, with the exception of 
the two transient magnetars \xte\, and \wes, and with the inclusion of 
recent \XMM\, observations of \cxo\, and \sgrd. The discussion of our 
findings is organized as follows. In \S\ref{xmmdisc}, we first concentrate 
on the results emerging from the fit of the \XMM\ data only, i.e. on the 
energy band below 10~keV. We report the discussion of our spectral 
results, a comparison with the analogous findings from the 1D RCS fits of 
\cite{nanda2}, and we discuss the cases of \ea\ and \uu, i.e. the only two 
sources for which a fit with the NTZ model is unsatisfactory. A search for 
correlations among the sources and the model parameters, in the same 
energy band, is reported in \$\ref{corr}. In \$\ref{intdisc} we then 
discuss the NTZ model in the contest of the interpretation of the whole 
SED up to $\sim$200~keV, basing this time on both \XMM\ and \INT\ data. 
The main limitations and caveats are then summariszed in \$\ref{cav}.

\subsection{The soft X-ray spectra: fits of the \XMM\ data}
\label{xmmdisc}

Our results on the fits of the \XMM\ data are reported in 
Tab.~\ref{tablexmmnonvar}, and 
Figs.~\ref{spectraxmmnonvar1},\ref{spectraxmmnonvar2} for sources without 
a significant spectral variation, and in Tables~\ref{tablexmmvar1}, 
\ref{tablexmmvar2} and Fig.\,\ref{spectraxmmvar} for sources that do 
exhibit spectral variation. When we restrict to the 1--10~keV band, we 
found that the NTZ model successfully reproduces the soft X-ray part of 
the spectrum of most of the sources (apart from \ea\ and \uu), without the 
need of additional components.  This represents a substantial improvement 
with respect to previous attempts to model magnetars quiescent emission in 
the same energy band with a simpler 1D RCS model \citep{nanda2}, where it 
was found that in a few cases a PL component was required, in addition to 
the RCS one, to provide an acceptable fit to the data below 10~keV.
In particular, this was the
case of few AXPs (including e.g. \rxs and \kes) and of   
\sgra, all them detected also above $\ga 20$~keV. In such
cases, we found that the 1D RCS 
component reproduces the
spectrum only up to 5--8\,keV. In order to match the data at the
highest \XMM\ energies, the contribution of a PL must be included. On 
the other hand, the slope of this additional PL is the same of that 
describing the \INT\ spectrum in the $\sim 20$--200~keV band. 
It is well possible that the mechanism responsible for the hard 
X-ray
emission provides a non negligible contribution in the soft X-ray
range. However, our finding is that the resonant Compton scattering model 
by
\cite{ntz1} correctly describes the data in the whole \XMM\ energy
range, i.e. up to 10~keV, for all the sources (and source states) 
reported in 
Tables~\ref{tablexmmnonvar},\ref{tablexmmvar1},\ref{tablexmmvar2}. The 
application of the NTZ model to the $\ga 20$~keV
emission from magnetars is discussed later on.  It is worth noticing
that the NTZ model has two free parameters less than the 1D RCS+PL
model used in \cite{nanda2}. Hence, besides self-consistency, it is
also more robust on a statistical ground. 

In all cases we found that $N_H$, as derived from the NTZ model, is
lower than (or consistent with) that inferred from the BB+PL fit, and
consistent with what derived from fitting the single X-ray edges 
(Durant \& van Kerkwijk 2006). The same was also true for the fits with the
1D RCS model presented in \cite{nanda2}. This is not surprising, since
the power-law usually fitted to magnetar spectra in the soft X-ray
range is well known to overestimate the column density.\footnote{This
  is because the fitting procedure tends to increase absorption
  (i.e. $N_H$) to counter the steep rise of the power-law at low
  energies, which eventually diverges as $E\to 0$.}  The surface
temperature we derived fitting the NTZ model is slightly higher than
the corresponding 1D RCS temperature and consistent with the BB
temperature in the BB+PL model. On the other hand, a quantitative
comparison between the values of $\beta_{bulk}$ and $\Delta \phi$,
i.e. the parameters that describe the magnetospheric currents in the
NTZ model, with the corresponding parameters of the 1D RCS model
($\beta_T$ and $\tau_{res}$) is more difficult, due to the different
assumptions about the currents velocity distribution and the magnetic
field topology. The 1D RCS assumes a plane parallel 
slab (i.e.  photons can only propagate along the slab 
normal, either towards or away from the star). 
Moreover, magnetospheric charges are assumed to have a
top-hat velocity distribution centered at zero and extending up to
a given temperature, $\pm kT_e$. To some extent, this scenario mimics a 
thermal, 1D, motion
(in which case $kT_e$ can be assimilated to a mean $e^-$ energy, i.e. to
the temperature of the 1D electron plasma). Since the $e^-$ velocity
distribution averages to zero, no bulk motion is accounted for: if we
only consider the frequency shift due to the $e^-$ motion then a
photon has the same probability to undergo up or down
scattering. Photon boosting by particle thermal motion in Thomson
limit may still occur, because the spatial variation of the magnetic
field is taken into account.  For a photon propagating from high to
low magnetic fields, multiple resonant cyclotron scattering will, on
average, up-scatter the transmitted radiation, giving rise to the
formation of an hard tail.  
Since, at variance with the 1D RCS model, our code
accounts for both bulk and thermal motion, one may expect to find a
value of $\beta_{bulk}$ systematically lower than the value of the
velocity parameter $\beta_T$ in the 1D RCS fits.  However, we find
that this is not always the case and the relation between the values
of $\beta_{bulk}$ and $\beta_T$ appears to be more complex. The NTZ
model does not explicitly contain the optical depth as a
parameter. However, $\Delta \phi$ and $\beta_{bulk}$ can be related to
the (average) scattering depth, as discussed in \cite{nanda2}. In
particular, the value of $\tau_{ave}$ can be read off from their
figure 10, simply by dividing the value of the depth corresponding to
a given $\Delta \phi$ by $\beta_{bulk}$. The values of $\tau_{ave}$ we
derive ($\sim 0.1$--1, as expected since the average number of
scatterings for a typical photon is of this order) are systematically
lower than those obtained for the 1D RCS $\tau_{res}$. This means that
the 1D RCS model is intrinsically less efficient in up-scattering the
seed photons. Its spectrum is, in fact, softer and it requires a larger
depth (and hence a larger number of scatterings). Moreover, the 
hardest sources require an additional PL to match the observed \XMM\  
spectra.
When observations of the same source at different epochs are
available, it is of interest to check if the best-fit values of the
model parameters change, since this can reveal how the physical
properties of the surface/magnetosphere evolve in time. In the case of
\ee\ all the parameters are compatible with being constant within the
errors, with the only 
exception of 
the normalization which appears to increase following the flux rise.  The
same holds for \sgra\, where there may be also an indication for a
decrease in the twist angle after the giant flare of December 27th
2004. Errors are however quite large and prevent any firm conclusion
at this stage.  The time behaviour of \1e\ is puzzling, since
$\beta_{bulk}$ increases while $\Delta\phi$ decreases as the flux, and
the model normalization, increases. In this source there is also a
quite robust evidence that the surface temperature increased with the
flux.

As we mentioned before, in the case of \ea\ and \uu\ we were unable to 
find a
satisfactory fit with the NTZ model. 
We noticed that, below 10~keV, these
two sources are characterized by a rather soft spectrum:  in the canonical
BB+PL decomposition the power law index is $\sim 4$ instead of $\sim2-3$
as in other magnetars. At the same time, the power law tail starts very
close to the energy at which the BB peaks. Such a spectral shape appears
difficult to explain in terms of upscattering of
soft photon, whatever the nature of the comptonization process  might be.
In fact, a PL component that starts close to the BB peak is a signature of a
full fledged comptonization in which case, however, it is expected to be
quite flat.
Conversely, a steep power law tail is typical of weak comptonization and departures
from the BB spectrum occur at energies beyond the peak. One possible
explanation is that the BB peak {\it appears} to be less prominent to the
observer because the region that emits the soft seed photons is not coming
completely into view. In \cite{ntz1}, we considered the effects of a
non-homogeneous surface temperature distribution, by examining  the case in which photons are
emitted by a single surface patch. In Fig.~\ref{onecap} we show the results of
our simulations in the case in which the emitting region is
confined to an equatorial strip (left panel) or to a polar cap (right panel).
The subdivision of the star surface, that of the
sky, the energy range and bin width are the same as that used in Nobili, 
Turolla \& Zane (2008a; see Fig~7 in that paper). 
The different curves show the
emerging spectrum, as viewed by an observer whose line of sight (LOS) 
makes an angle
$\Theta_s$ with the spin axis and for different values of the observing
longitude, $\Phi_s=20^\circ,\,
140^\circ,\, 220^\circ$.
These three values correspond to having the
emitting surface patch in full view (seen nearly face on), partially in
view and almost screened by the star. As it can be seen, when the emitting
patch is in full view the observed spectrum consists of a well visible
thermal component and an extended power-law-like tail.
On the other hand, if the emitting region is not
directly visible, no contribution from the primary blackbody photons is
present: the spectrum, which is made up only by those photons which after
scattering propagate ``backwards'', has a  depressed thermal peak and
a much more distinct non-thermal shape. In particular, in the
examples of Fig.~\ref{onecap},   the scattering efficiency is not very
high and curves corresponding to  $\Phi_s = 220^\circ$ have a steep
power-law tail with photon index $\sim 4$, as observed in \ea\ and \uu.
Although
a quantitative fit of the spectra of these two sources including different
thermal maps in the code would be
unfeasible (because it would introduce too many degrees of freedom), it is
tempting to speculate that the
peculiar spectrum of \ea\ and \uu\  is due to a strong inhomogeneity in the
surface temperature distribution, with the hotter region almost antipodal
with
respect to the observer. This is also compatible with the fact that these
two sources have a rather low pulsed fraction with respect to other 
magnetars.
On the other hand, the double peaked pulse profile of \ea\ is difficult to
explain in this picture. Another possibility is that the phase
average spectrum appears to be quite soft because it reflects the
contribution of different components. As discussed by \cite{wo04} 
and \cite{rea07}, 
both these sources exhibit a strong spectral variation with spin phase. In 
the case 
of \uu\ the  spectrum
switches from being very hard to very soft within a 0.1-wide phase 
interval and \XMM\  data
reveals a  discontinuity, between 2 keV and 3
keV, which can be interpreted as a curved
component and is  most appearent within phase interval 0.7-0.9  
\citep{dh08a}. In order
to assess
this scenario, a more detailed
investigation of the pulse resolved spectra is necessary; this requires
the introduction of the viewing geometry in our fits  and will be
presented in  a forthcoming paper.

\subsection{Correlations}
\label{corr}

We run a number of Spearman's rank correlation tests, in order to look for
possible links among the observed properties of the sources in our sample
and the model parameters. In particular, we checked for correlations
between the 1--10 keV luminosity, $L_{1-10\, {\rm keV}}$ or the spin-down
value of the magnetic field, $B$, and each of the NTZ parameters, $kT$,
$\beta_{bulk}$, and $\Delta\phi$. The values of the parameters are 
those obtained from the fit of the \XMM\ data only 
(Tables~\ref{tablexmmnonvar}, ~\ref{tablexmmvar1}, ~\ref{tablexmmvar2}). 
The source 
distances and the values 
of $B$ have been taken from the compilation in \cite{nanda2}. The only
significant correlations which emerged are those between $\beta_{bulk}$
and $L_{1-10\, {\rm keV}}$ and $\beta_{bulk}$ and $B$. Both show a
deviation from the null hypothesis probability of $\sim 93\%$. Although
the significance level is lower, $81\%$, a correlation between $kT$ and
$L_{1-10\, {\rm keV}}$ seems also to be present. The correlation between
$L_{1-10\, {\rm keV}}$ and $\beta_{bulk}$, which is direct, can be
explained taking into account that an increase of $\beta_{bulk}$ results
in a larger energy gain of the photons per scattering and hence in a
hardening of the spectrum, which translates into a higher luminosity. A
similar argument applies to the $L_{1-10\, {\rm keV}}$--$kT$ correlation,
which is again direct. An increase of the surface temperature implies an
increase of the flux of primary photons and again of the observed
luminosity. The $\beta_{bulk}$--$B$ correlation mirrors that between
$\beta_{T}$ and $B$ reported in \cite{nanda2}, and, as discussed there, is
not of immediate interpretation.

As discussed in \cite{ntz1}, the modulation of the X-ray flux may
be due to the anisotropy of the magnetospheric charge density
(which is lower along the poles), to a patchy surface temperature
distribution, or to a combination of both effects. 
As mentioned in \S\ref{mod}, in  NTZ models the normalization constant 
(norm) divided by $kT^3$ is proportional to the emitting area on the star 
surface. 
To check if
the observed modulation is related (also) to the presence of a
hotter region on the surface, we looked for a correlation between
the pulsed fraction and the ratio norm$\times D^2/kT^3$, where $D$
is the source distance. The pulsed fraction is defined as the 
semiamplitude of the sinusoidal function that best-fits the lightcurve, in 
the same energy 
range. 
We run the test on the entire sample of
sources, excluding the first observation of \1e, for which only an
upper limit of the pulsed fraction is available. No significant
correlation was found. However, when the sample is restricted to
AXPs only, a negative correlation (i.e. the pulsed fraction
increases when the area decreases) emerges at the $91\%$
confidence level. To verify this, we run again the test including
the data from a set of five more recent observations of \1e\
(Bernardini et al., in preparation) and found that the correlation
is still present, although the significance level slightly
decreases to $74\%$ (see Fig.~\ref{pf-axp}). Taken face value, this
results points towards a localized emitting area in AXPs while the
surface temperature would be more uniform in SGRs, where the
modulation is produced mainly by scatterings in the magnetosphere.
We note that the presence of a (time varying) hot spot has been
reported in some transient AXPs, but the existence of a general
pattern as the one discussed above needs a larger sample to be
confirmed.

Finally, we checked if there is a correlation between $\Delta\phi$ and
$\beta_{bulk}$. The motivation for this is that both parameters are
responsible for the formation of the high-energy tail. This may
introduce a redundancy in the model parameters: in principle the fit
might be not unique, since quite similar spectra may be obtained with
different combinations of $\Delta\phi$ and $\beta_{bulk}$ (e.g. low
$\beta_{bulk}$ and large $\Delta\phi$ or the opposite). We did not
find any significant correlation between these two quantities.

\subsection{The broadband spectrum in the 1-100~keV range.}
\label{intdisc}

As discussed in \S\ref{appl}, a spectral decomposition of the kind
NTZ+PL successfully reproduces the whole spectrum of \rxs, \kes, and
\sgrb, up to 200~keV. In this case, however, the best fit
parameters of the NTZ model differ from those found from the
fit of \XMM\ data up to 10~keV. In particular, while the temperature
of the seed blackbody is almost unchanged, the value of $\beta_{bulk}$
is always considerably reduced (from 0.3-0.5 down to 0.1-0.2) and the
twist angle less constrained. This indicates that in the broadband fit
most of the hardening is accounted for by the additional PL component
that, in the case of \kes\ and \sgrb, starts to dominate the spectrum
at energies as low as $\sim$3--4~keV. Of course, this kind of spectral
decomposition is certainly possible and may mimic a scenario in which
the hard X-ray and soft X-ray emissions are due to two separate
processes: for instance soft $\gamma$-rays may be produced in a
twisted magnetosphere by thermal bremsstrahlung emission from the
surface region heated by returning currents, or synchrotron emission
from pairs created higher up ($\sim$ 100 km) in the magnetosphere
\citep{tb05}.

In principle, however, spectra produced by resonant
cyclotron upscattering of soft photons are expected to develop power
law tails that extend up to much higher energies
\citep{baring,ntz1}. Therefore, it is well possible that our model can
consistently explain the whole spectral energy distribution in the
1-200~keV range.  If this is the case, the slope of the hard power law
tail would be mainly dictated by the properties of the magnetospheric
electrons responsible for the upscattering, i.e. their density and
velocity distribution. Observationally the hard X-ray properties of
AXPs and SGRs are quite different: while in the range 4-200~keV the
spectra of SGRs are dominated by the non-terhmal component and, in the
canonical model, well reproduced by a single unbroken PL, AXPs show a
sort of turn over and have a non-thermal tail which is initially
softer, up to $\sim 5-8$~keV, and then flattens at higher energies. In
this picture, \kes\, seems an exception: its non-thermal emission is
well reproduced by a single PL and is therefore more SGR-like
\cite[see][and references therein]{nanda2}. These properties suggest
that, within the resonant Compton scattering scenario, AXPs spectra
require different electron populations. 
On the other hand, a single electron distribution as for
instance the one we used in the 3D computations presented here, may
account for the broadband emission of SGRs or SGR-like sources.

In order to test this possibility, we can not use directly our XSPEC table 
of models, since it has been computed assuming that magnetic
scattering is conservative in the electron frame and, for
self-consistency, spectra were truncated at 15~keV (see \S~\ref{mod}
for a discussion). Instead, 
we re-computed the model that
best-fits the 1-10~keV spectrum of \sgrb (which parameters are reported in 
Table~\ref{tablexmmnonvar}), 
this time by extending
the computation to the whole 1-200~keV energy range and using a fully
relativistic version of the Monte Carlo code which incorporates the
complete QED scattering cross section \citep{ntz2}. We also repeated
the computation by using slightly different sets of parameters, all
within $3\sigma$ from their best-fitting values (errors at $1\sigma$
are reported in Table~\ref{tablexmmnonvar}). As expected, these new
spectra differ from those computed in the Thomson limit only at very
high energies ($>100$~keV).  Finally, we added to each curve a free
normalization constant in the \INT\ range, to account for
inter-calibration uncertainties between the two satellites. We then
fitted the \INT\ data by leaving only the intercalibration constant as
a free parameter. Results are reported in Fig.~\ref{intfit}, where the
two curves correspond to: i) the best fitting parameters reported in
Table~\ref{tablexmmnonvar} and ii) to a different model in
which $\beta_{bulk}$, $kT$ and $\Delta \phi$ have been increased up to
their upper limits within the $3\sigma$ confidence range from the
\XMM\ fit. We
found that in both cases the fit to \INT\ data is excellent (reduced
$\chi^2=1.14$ and 1.07, respectively, for 7 degrees of freedom), but
the two model normalizations, in the \XMM\ and \INT\ ranges, differ by
a factor 8.8 and 2.8 respectively, too large to be attributed to
intercalibrations uncertainties. Not surprisingly, similar large factors 
are also found 
when trying to fit the \XMM\ and \INT\ data in the range 6-200~keV with 
a simple power law model. On the other hand, this 
simple 
test proves that the spectral slope of our model in the 20-200~keV range
is the same as that of the \INT\ data. 
Similar considerations hold in the case of \kes. 
Although a conclusive
answer requires a direct fitting of the combined \XMM\ and \INT\ data 
with a self-consistent Compton model, which is beyond the 
scope of this paper, we regard this preliminary finding as promising. 

\subsection{Caveats and future developments} 
\label{cav} 

We caveat that the models presented here are based on a number of
simplifying assumptions. First, they are based on a globally
twisted dipole model, that only gives an idealized representation of
the magnetic field topology.  There are now both theoretical
\citep{bel09} and observational (a certain degree of hysteresis in the
long-term evolution of SGR 1806-20, \citealt{wo07}; the long-term
evolution of the thermal component of the transient AXP XTE J1810-197,
\citealt{perna, bern08}) motivations in favour of a picture in which
the twist affects only a limited portion the magnetosphere, typically
the polar region. 
Furthermore, phase resolved spectroscopy of the two
AXPs 1RXS J1708-4009 and 4U 0142+61 in the \INT\ energy range
\citep{dh08a, dh08b} shows dramatic spectral changes with the spin phase. 
It
has been recently suggested that a resonant scattering model in which
the magnetic field is locally twisted may catch the essential features
of this behaviour \citep{pav08}.

A further point is the nature and velocity distribution of the
magnetospheric charges. The NTZ model assumes the presence of mildly
relativistic electrons, moving at constant velocity (which is a model
parameter) along the closed field lines. Currents flowing in the
magnetosphere of a magnetar have been investigated by \cite{beltho07},
who concluded that the magnetosphere is populated by pairs with a
Lorentz factor $\gamma\approx 1000$. However, more recent calculations
(Nobili, Turolla \& Zane, in preparation) show that in the region
where $B\ga 2 B_Q$ Compton losses efficiently slow down electrons,
limiting pair creation to a small region close to the star surface,
and allowing for the presence of mildly relativistic particles along
much of the circuit.

Finally, angle of view effects have not been accounted for: the
emerging spectrum is simply computed by integrating over the whole sky
at infinity. This corresponds to the case in which the star is an
aligned rotator, i.e the spin and magnetic axes coincide.  As
discussed in \cite{ntz1}, in order to treat the more general case in
which the spin and magnetic axes are not aligned, we also produced a
second XSPEC {\tt atable} model by introducing two angles, $\chi$ and
$\xi$, which give, respectively, the inclination of the LOS and of the
dipole axis with respect to the star spin axis. This allows us to take
into account for the star rotation and hence derive pulse shapes and
phase-resolved spectroscopy. The spectral model, {\tt ntzang.mod}, has
six free parameters ($\beta_{bulk}$, $\Delta \phi $, $\log kT$,
$\chi$, $\xi$ plus a normalization constant), i.e. two more than the
one used here. Given that the NTZ model produces a very good fit for
the phase-averaged spectra, the inclusion of two further parameters
(i.e. the two angles) is not statistically required, as we tested, and
will leave $\chi$ and $\xi$ unconstrained. On the other hand, having
the possibility to infer the viewing geometry may be useful when
fitting different outburst states in transient AXPs or when combining
information that can be obtained by fitting simultaneously
phase-resolved spectra, or independently from the study of the pulse
profile. Further work in this direction is under way and will be
presented in separate papers.

\section{Conclusions}
\label{conc}

By considering a large sample of magnetars, we found that resonant
Compton upscattering by a population of mildly relativistic electrons
($\beta_{bulk} \sim 0.1--0.7$) can reproduce the pulse averaged spectra
in the range 1--10~keV. At variance with the 1D RCS model adopted in
\cite{nanda2}, the approach used in the present investigation
consistently accounts for the bulk motion of magnetospheric electrons
which results in a more efficient comptonization of seed thermal
photons. This has two main consequences: i) the required values of the
optical depth are lower that those found using the 1D RCS, and ii) NTZ
spectra, being intrinsically harder, successfully reproduce also the
SGRs power-law tail below 10~keV. 
We found a significant correlation 
between
the 1--10 keV source luminosity and both $\beta_{bulk}$ and $kT$.
This is indeed expected in the resonant cyclotron scattering model and
further supports this scenario to explain the high energy emission
from magnetars. Moreover, when restricting to AXPs only, we find hints
for a negative correlation between the pulsed fraction and the
emitting area. If confirmed, this suggests the presence of a strong
thermal gradient on the star surface, which may also be responsible
for the slightly different spectrum of \uu\ and \ea, the only two
sources for which we could not find a satisfactory fit. Anisotropic
surface thermal distributions may arise in the presence of large
crustal magnetic fields, as expected in magnetars, because heat is
preferentially transferred along the field, resulting in small and
confined hot caps \cite[e.g.][]{ge06,pmg09}.  Moreover, the magnetar
surface is also expected to be further heated during bursting activity
or by returning currents; in both cases the heat deposition can be
substantially anisotropic.

Finally, our conclusions regarding the modelling of the whole SED 
distribution up to $\sim200$~keV are still compatible with various 
scenarios. In the case of \kes, \rxs, and \sgrb, a double component NTZ+PL 
model fits the combined \XMM\ and \INT\ data,  
but it requires a NTZ spectrum quite soft and most of the emission 
dominated by the additional PL. This is compatible with a scenario in 
which soft X-ray and hard X-ray emission are ascribed to independent 
components. On the other hand, there is also the possibility that models 
of resonant upscattering can successfully describe the whole SED 
including the hard X-ray tail 
(eventually by invoking more than one electron population for 
sources with a spectral turnover at high energies). 
A conclusive answer requires a direct fitting of the combined \XMM\ and 
\INT\ data, with all parameters left free and a new XSPEC table of models 
computed by using the QED cross sections. Further work on these
topics is in preparation.

\section*{Acknowledgments}
SZ acknowledges STFC for support through an Advanced Fellowship. NR is
supported by an NWO Veni Fellowship.  RT and LN are partially
supported by INAF-ASI through grant AAE-I/088/06/0. 
We thank D. G\"otz for providing the INTEGRAL spectra used in this
paper, A. Tiengo and P. Esposito for the data of \cxo\ and \sgrd, 
F. Bernardini and G.L. Israel for the additional
\1e\ data shown in Fig.\ref{pf-axp}.

%%%%%%%%%%%%%%%%% tables and figures  %%%%%%%%%%%%%%%%%%%%

%%%%%%%%%%%%%%%%% sorgenti con 1 singolo stato, xmm only %%%%%%%%%%%%

\begin{table*}
\setlength{\tabcolsep}{0.02in}
\centering
\begin{tabular}{lccccc}
\hline
\hline
  & \multicolumn{1}{c}{1RXS\,J1708--4009$^*$} &
\multicolumn{1}{c}{1E\,1841--045} &
\multicolumn{1}{c}{\sgrb} &
\multicolumn{1}{c}{\cxo} &
\multicolumn{1}{c}{\sgrd}
\\
\hline
 \multicolumn{1}{l}{Parameters} &  &  &    \\
\hline
N$_{H}$ &  1.45$\pm$0.08 & 2.5$\pm$0.1 & 3.74$\pm$0.15 &
$<0.05$ & 12$\pm$1 \\
 & & &  & & \\
%log kT (keV) &  -0.32$\pm$0.08 & -0.30$\pm$0.05
%&
%-0.35$\pm$0.04 &
%-0.46$\pm$0.02 & -0.20$\pm$0.07 \\
kT (keV) &  0.47$\pm0.09$  & 0.50$\pm0.06$
& 0.45$\pm$0.04 & 0.35 $\pm$0.02 & 0.64$\pm0.11$ \\
$\beta_{bulk}$ & 0.34$\pm$0.04 & 0.43$\pm$0.05 & 0.46$\pm$0.05
& 0.18$\pm$0.03 & 0.7$\pm$0.1  \\
$\Delta \phi$ &  0.49$\pm$0.15 & 0.47$\pm$0.04  &  0.45$\pm$0.03
&   1.7$\pm$0.8 & 1.3$\pm$0.4 \\
NTZ~norm &  0.35$\pm$0.04 & 0.20$\pm$0.02 &  0.08$\pm$0.02
& 0.004$\pm$0.001 & 0.005$\pm$0.001 \\
 & & &  & & \\
Flux (1--10\,keV) & $(2.6\pm0.2)\times10^{-11}$
& $(2.1\pm0.2)\times10^{-11}$
& $(3.9\pm0.3)\times10^{-12}$  & $(3.3\pm 0.2) \times10^{-13}$
&$(3.3\pm0.2)\times10^{-13}$ \\
 & & & & &  \\
$\chi^2_{\nu}$ (dof) & 0.97 (197)   &  1.04 (152) &  0.99 (135)
&   1.21 (101) &  1.16 (81)   \\
\hline
\hline
\end{tabular}
\caption{Spectral Parameters of \rxs, \kes,  \sgrb,
\cxo, and \sgrd\ obtained by
fitting the \XMM\ data with an NTZ  model. The fit has been restricted to
the 1--10\,keV range, with the exception of \cxo\ and \sgrd\ for which 
we used
the 0.1--10\,keV  and 2--10\,keV band, respectively. Errors are at 
1$\sigma$
confidence level,
reported fluxes are absorbed and in units of \ergscm2. N$_{H}$ is in
  units of $10^{22}$\,cm$^{-2}$ and we assumed solar abundances
from Lodders (2003).
See also Fig.\,\ref{spectraxmmnonvar1},
\ref{spectraxmmnonvar2} and
  \S\,\ref{appl} for details.
\newline $^{*}$: source  slightly variable
in
  flux and spectrum, see text for details. }
\label{tablexmmnonvar}
\end{table*}

%%%%%%%%%%%%%%%%% sorgenti con piu' stati, xmm only %%%%%%%%%%%%

%%%%%%%%%%%%%%%%%%%%%%%%% 1E 1547  and 1048 %%%%%%%%%%%%%%%%%%%%%%%%%%%%%%%%%%%%%
\begin{table*}
\setlength{\tabcolsep}{0.02in}
\centering
\begin{tabular}{lccccc}
\hline
\hline
    & \multicolumn{2}{c}{\1e}  & \multicolumn{3}{c}{\ee} \\
\hline
  Parameters  & \multicolumn{1}{c}{2006} &  \multicolumn{1}{c}{2007} &
\multicolumn{1}{c}{2003} &  \multicolumn{1}{c}{2005} & \multicolumn{1}{c}{2007} \\
\hline
N$_{H}$ & \multicolumn{2}{c}{4.6$\pm$0.13} & \multicolumn{3}{c}{0.66$\pm$0.02} \\
 & & & & & \\
%log kT (keV) &  -0.42$\pm$0.01 & -0.25$\pm$0.01   &  -0.22$\pm$0.04 &
%-0.25$\pm$0.06 & -0.16$\pm$0.01\\
kT (keV)
&  0.38$\pm$0.01 & 0.56$\pm$0.01 & 0.60$\pm 0.06$ &
0.56$\pm 0.08$
& 0.69$\pm$0.02\\
$\beta_{bulk}$ & 0.15$\pm$0.05 & 0.4$\pm$0.1  &  0.14$\pm$0.03 &
0.16$\pm$0.07 & 0.13$\pm$0.04\\
$\Delta \phi$ &   1.14$\pm$0.08 & 0.4$\pm$0.1  & 1.9$\pm$0.2 &
2.0$\pm$0.2 &  1.9$\pm$0.1\\
NTZ~norm &   0.02$\pm$0.01 & 0.082$\pm$0.005 &  0.10$\pm$0.03 &  0.07$\pm$0.01 &  0.21$\pm$0.01\\
  & & & & \\
Flux (1--10\,keV) &  $(3.3\pm0.2)\times10^{-13}$ &
$(3.0\pm0.2)\times10^{-12}$ &  $(1.1\pm0.2)\times10^{-11}$ &
$(8.3\pm0.4)\times10^{-12}$  &  $(3.0\pm0.2)\times10^{-11}$\\
 & & & & &  \\
$\chi^2_{\nu}$ (dof) &   \multicolumn{2}{c}{1.11 (164)} & \multicolumn{3}{c}{1.22 (515)}  \\
\hline
\hline
\end{tabular}
\caption{Spectral Parameters of \1e\, and \ee in different emission states, obtained by
fitting the \XMM\, observations with an NTZ model.
The fit has been restricted to
the 1--10\,keV range. Errors are at 1$\sigma$
  confidence level,
reported fluxes are absorbed and in units of \ergscm2. N$_{H}$ is in
  units of $10^{22}$\,cm$^{-2}$ and we assumed solar abundances from
  Lodders (2003). See also
Fig.\,\ref{spectraxmmvar}.}
\label{tablexmmvar1}
\end{table*}

%%%%%%%%%%%%%%%%%%%%%%%%% SGR1806 %%%%%%%%%%%%%%%%%%%%%%%%%%%%%%%%%%%%%
\begin{table*}
\setlength{\tabcolsep}{0.02in}
\centering
\begin{tabular}{lccc}
\hline
\hline
    & \multicolumn{3}{c}{\sgra} \\
\hline
  Parameters   & \multicolumn{1}{c}{2003} &  \multicolumn{1}{c}{2004} &  \multicolumn{1}{c}{2005} \\
\hline
N$_{H}$ &  \multicolumn{3}{c}{9.1$\pm$0.5} \\
 & & &   \\
%log kT (keV) & -0.16$\pm$0.06 & -0.10$\pm$0.05 & -0.13$\pm$0.08 \\
kT (keV) & 0.69$\pm 0.10$ &
0.79$\pm 0.10$ & 0.74$\pm 0.15$\\
$\beta_{bulk}$ &    0.48$\pm$0.05  & 0.52$\pm$0.05 & 0.51$\pm$0.05  \\
$\Delta \phi$ &    1.8$\pm$0.8 & 1.7$\pm$0.7 &  1.0$\pm$0.9    \\
NTZ~norm &   0.11$\pm$0.05 & 0.23$\pm$0.05    &    0.12$\pm$0.08  \\
  & & & \\
Flux (1--10\,keV) &  $(1.2\pm0.1) \times 10^{-11}$ &  $(2.6\pm0.2)\times
10^{-11}$  &
$(1.3\pm0.2)\times 10^{-11}$ \\
 & & &   \\
$\chi^2_{\nu}$ (dof) &  \multicolumn{3}{c}{0.98 (288)} \\
\hline
\hline
\end{tabular}
\caption{Spectral Parameters of \sgra\, in different emission states, obtained by
fitting the \XMM\, observations with an NTZ model.
The three datasets have been taken before and after the giant flare of 2004 December 27
\citep{hu05, pa05}, with the intermediate observation taken about two
months before the event.  
The fit has been restricted to
the 1--10\,keV range. Errors are at 1$\sigma$
  confidence level,
reported fluxes are absorbed and in units of  \ergscm2. N$_{H}$ is in
  units of $10^{22}$\,cm$^{-2}$ and we assumed solar abundances from
  Lodders (2003). See also Fig.\,\ref{spectraxmmvar}.}
\label{tablexmmvar2}
\end{table*}

%%%%%%%%%%%%%%%%% sorgenti con integral aggiunto %%%%%%%%%%%%

\begin{table*}
\setlength{\tabcolsep}{0.02in}
\centering
\begin{tabular}{lcccc}
\hline
\hline
  & \multicolumn{1}{c}{1RXS\,J1708--4009$^*$} &
\multicolumn{1}{c}{1E\,1841--045} & \multicolumn{1}{c}{\sgrb}\\
\hline
 \multicolumn{1}{l}{Parameters} & NTZ+PL & NTZ+PL & NTZ+PL \\
\hline
N$_{H}$ & 1.67$\pm$0.03 & 2.5$\pm$0.1 & 3.9$\pm$0.2\\
 & & &  \\
%log kT (keV) & -0.41$\pm$0.04  &  -0.30$\pm$0.05 &  -0.37$\pm$0.06\\
kT (keV) & 0.39$\pm 0.04$ &  0.50$\pm 0.06$ &
0.43 $\pm0.06$\\
$\beta_{bulk}$ &    0.20$\pm$0.05 & 0.17$\pm$0.06 &
0.10$\pm$0.08 \\
$\Delta \phi$ & 1.90$\pm$0.04 & $>$ 0.72 & $>$ 0.38 \\
NTZ~norm & 0.42$\pm$0.04 & 0.28$\pm$0.03 & 0.06$\pm$0.01\\
 & & &  \\
$\Gamma$    & 0.96$\pm$0.07 &  1.46$\pm$0.08 &   1.87$\pm$0.09\\
PL~norm &   $(2.4\pm0.1) \ \times10^{-4}$ &  $(2.1\pm0.1)
\ \times10^{-3}$ &  $(1.1\pm0.1) \ \times10^{-3}$ \\
 & & &  \\
Flux (1--10\,keV) &  $(2.6\pm0.2)\times10^{-11}$
& $(2.1\pm0.2)\times10^{-11}$ & $(3.9\pm0.3)\times10^{-12}$ \\
Flux (1--200\,keV) &   $(1.1\pm0.2)\times10^{-10}$ &
 $(1.1\pm0.2)\times10^{-10}$ &   $(1.5\pm0.3)\times10^{-11}$ \\
 & & &  \\
$\chi^2_{\nu}$ (dof) & 1.05 (205)&   1.12 (158) & 1.02 (140)\\
\hline
\hline
\end{tabular}
\caption{Spectral Parameters of \rxs, \kes, and \sgrb\, obtained by
fitting the $\sim$1--200\,keV \XMM\, and \INT\, data with an NTZ+PL model.
A constant function has been included to the modelling to take into
  account for intercalibration uncertainties (always $<10$\%).
Errors are at 1$\sigma$ confidence level,
reported fluxes are absorbed and in units of \ergscm2. N$_{H}$ is in
  units of $10^{22}$\,cm$^{-2}$ and we assumed solar abundances
from Lodders (2003). See also Fig.\,\ref{spectraxmmintegral} and
  \S\,\ref{appl} for details.  \newline $^{*}$: source  slightly variable
in
  flux and spectrum, see text for details. }
\label{tableintegral}
\end{table*}

%%%%%%%%%%%%%%%%%%%%%%%%%%%%%%%%%%%%%%%%%%%%%%%%%%%%%%%%%%%%%%%%%%%%%%%

\newpage

\begin{figure*}
\centering{
\hspace{0.1cm}
\vbox{
\hbox{
\psfig{figure=1547_all_ntz_spec.ps,height=4.5cm,angle=270,width=4.5cm}
\psfig{figure=1048_all_ntz_spec.ps,height=4.5cm,angle=270,width=4.5cm}
\psfig{figure=1806_ntz_nopl_spec.ps,height=4.5cm,angle=270,width=4.5cm}}
\hbox{
\psfig{figure=1547_all_ntz_ufs.ps,height=4.5cm,angle=270,width=4.5cm}
\psfig{figure=1048_all_ntz_ufs.ps,height=4.5cm,angle=270,width=4.5cm}
\psfig{figure=1806_ntz_nopl_ufs.ps,height=4.5cm,angle=270,width=4.5cm}}}}
\caption{Same as in Fig.~\ref{spectraxmmnonvar1} for
the AXPs \1e, \ee\, and for \sgra. See
  Tables\,\ref{tablexmmvar1}, \ref{tablexmmvar2} for details.}
\label{spectraxmmvar}
\end{figure*}

%%%%%%%%%%%%%%%%%%%%%%%%%%%%%%%%%%%%%%%%%%%%%%%%%%%%%%%%%%%%%%%%%%%%%%%

\begin{figure*}
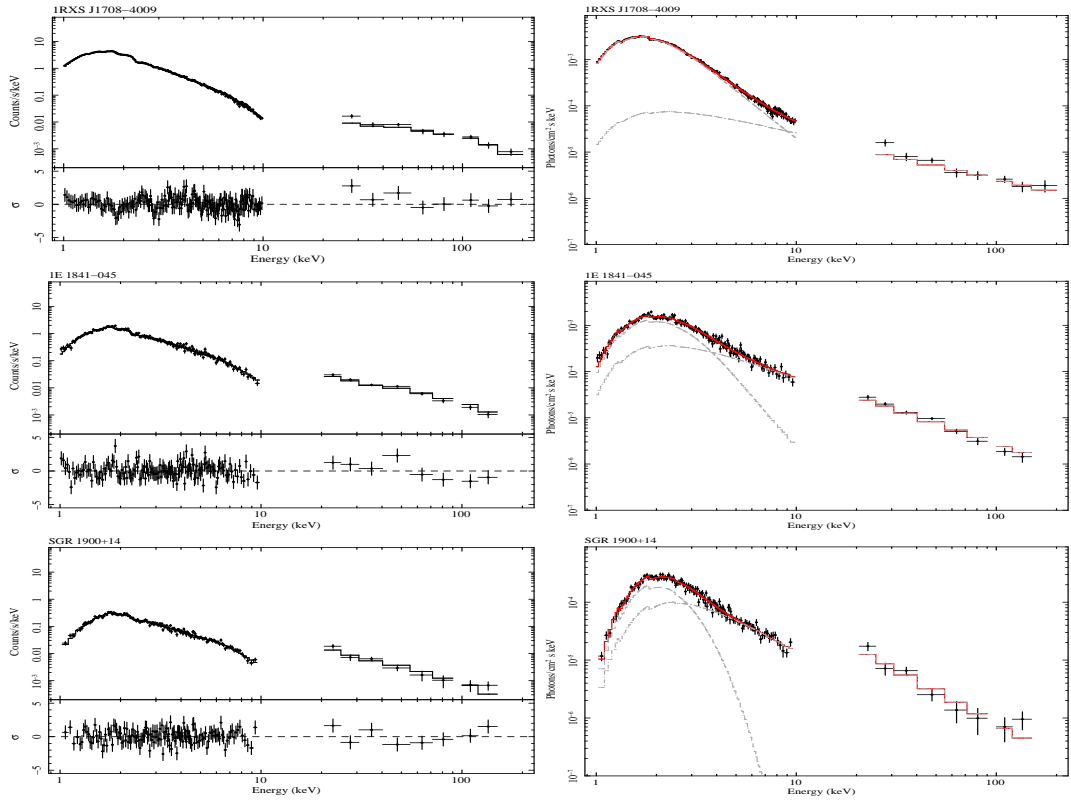

\centering{
\hspace{0.1cm}
\vbox{
\hbox{
\psfig{figure=1708_ntzpl_spec.ps,height=3.5cm,angle=270,width=7cm}
\psfig{figure=1708_ntzpl_ufs.ps,height=3.5cm,angle=270,width=7cm}}
\hbox{
\psfig{figure=1841_ntzpl_spec.ps,height=3.5cm,angle=270,width=7cm}
\psfig{figure=1841_ntzpl_ufs.ps,height=3.5cm,angle=270,width=7cm}}
\hbox{
\psfig{figure=1900_ntzpl_spec.ps,height=3.5cm,angle=270,width=7cm}
\psfig{figure=1900_ntzpl_ufs.ps,height=3.5cm,angle=270,width=7cm}}}}
\caption{\rxs, \kes, and \sgrb: left column shows the spectra in
  Count/s/keV while in the right
column we report the photon density plots of the modelling with the NTZ+PL 
model.
Both \XMM\  and \INT\  data have been used in the fitting.
See
  Table\,\ref{tableintegral} and
\S\,\ref{appl} for
  details. \label{spectraxmmintegral}}
\end{figure*}

%%%%%%%%%%%%%%%%%%%%%%%%%%%%%%%%%%%%%%%%%%%%%%%%%%%%%%%%%%%%%%%%%%%%%%%

%%%%%%%%%%%%%%%%%%%%%%%%%%%%%%%%%%%%%%%%%%%%%%%%%%%%%%%%%%%%%%%%%%%%%%%%%%%%%%%%%%%

\begin{figure*} \centering{ \hspace{0.1cm} \vbox{ \hbox{
\psfig{figure=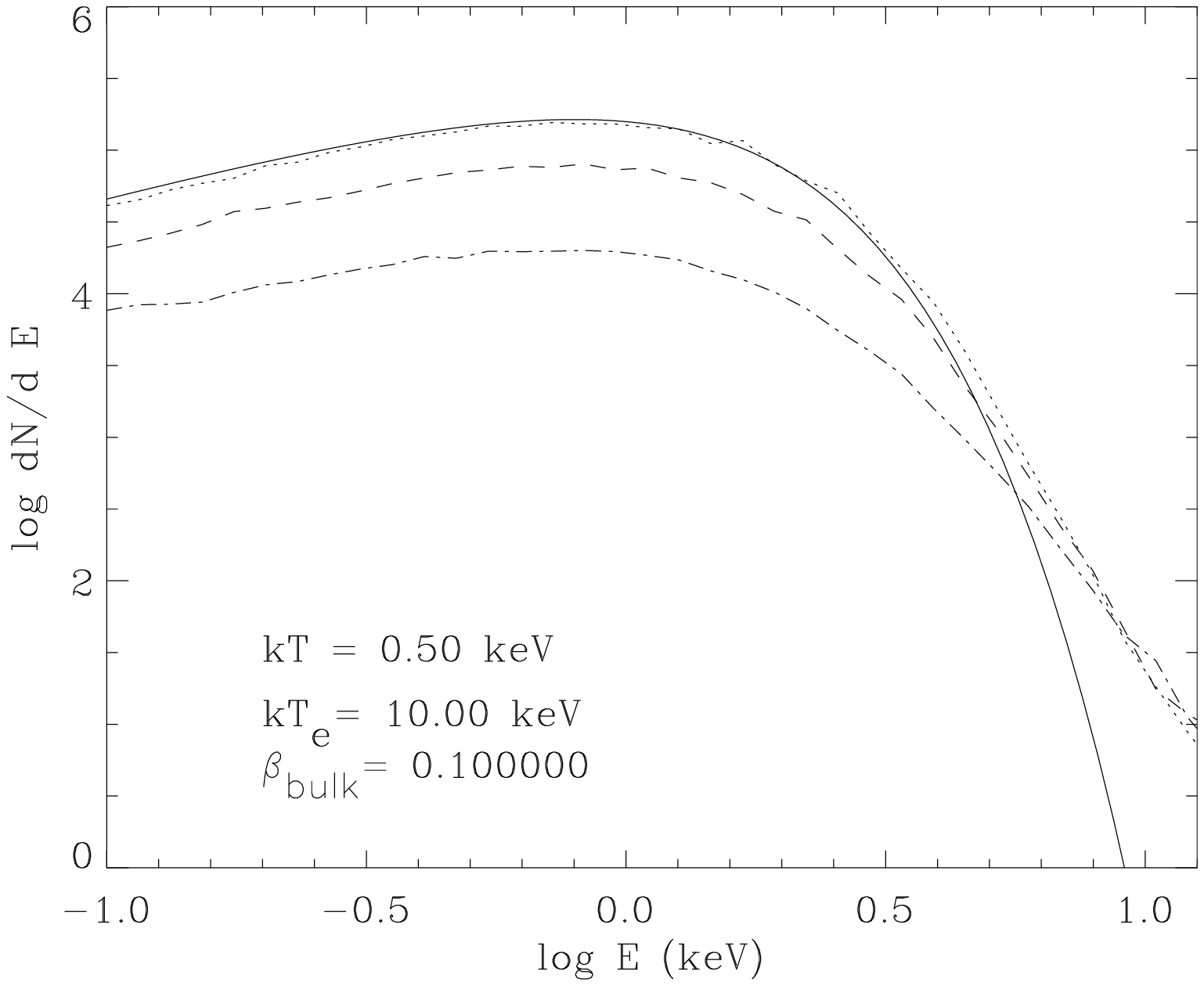,height=6.5cm}
\psfig{figure=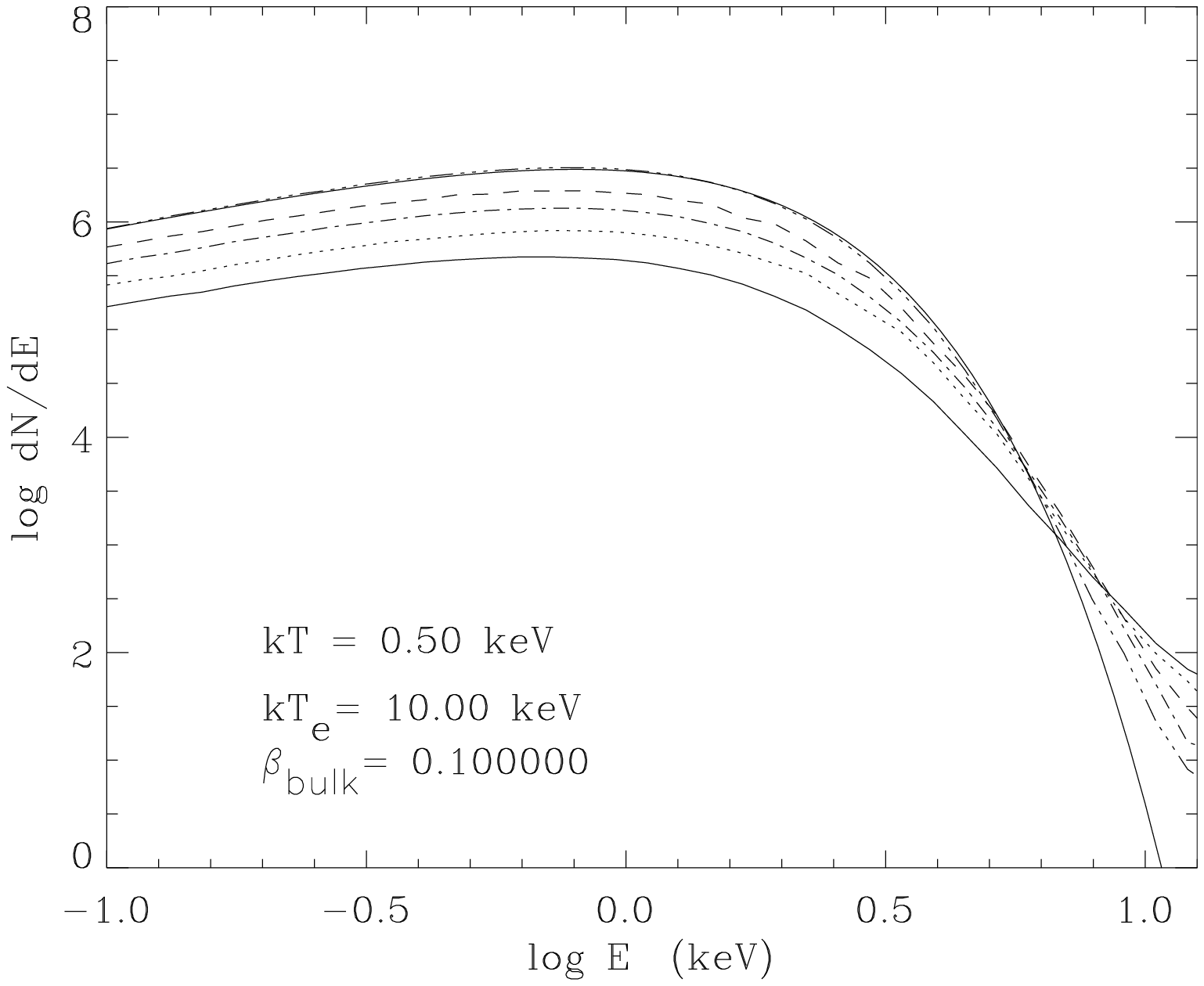,height=6.5cm}}}}
\caption{ Spectrum from a single
emitting zone on the star surface, computed as in Nobili, Turolla \& Zane
(2008a, see Fig.7 in that paper and text for details). Left: The emitting
area is the equatorial strip $0\leq\cos\Theta\leq 0.25$,
$0\leq\Phi\leq\pi/2$. The LOS is at $\Theta_s = 90^\circ$ and $\Phi_s =
20^\circ$ (dotted line), $\Phi_s = 140^\circ$ (dashed line) and $\Phi_s =
220^\circ$ (dash-dotted line). The solid line represents the seed
blackbody. Right:  The emitting area is the polar cap
$0.75\leq\cos\Theta\leq 1$, $0\leq\Phi\leq 2\pi$. The LOS is at $\Theta_s
= 19.5^\circ$ (dash-triple dotted line), $\Theta_s = 63.5^\circ$ (dashed
line), $\Theta_s = 90^\circ$ (dash-dotted line), $\Theta_s = 123.5^\circ$
(dotted line) and $\Theta_s = 160.5^\circ$ (lower solid line). The upper
solid line is the seed blackbody spectrum. } \label{onecap} \end{figure*}

\begin{figure*} \centering{  \hspace{0.1cm} \vbox{
\psfig{figure=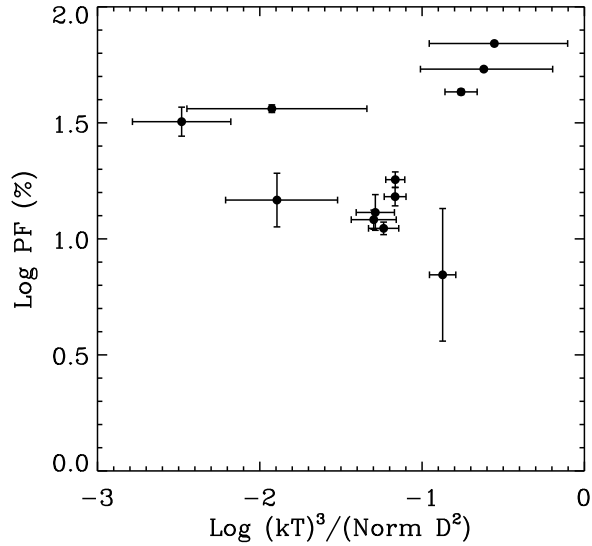,height=8.cm}}}
\caption{The pulsed fraction vs a measure of the surface emitting area for
the AXPs in our sample; see text for details. Five more recent
observations of \1e\  have been also included in this plot (Bernardini et
al., in preparation).
}
\label{pf-axp} \end{figure*}

\begin{figure*} \centering{  \hspace{0.1cm} \vbox{
\psfig{figure=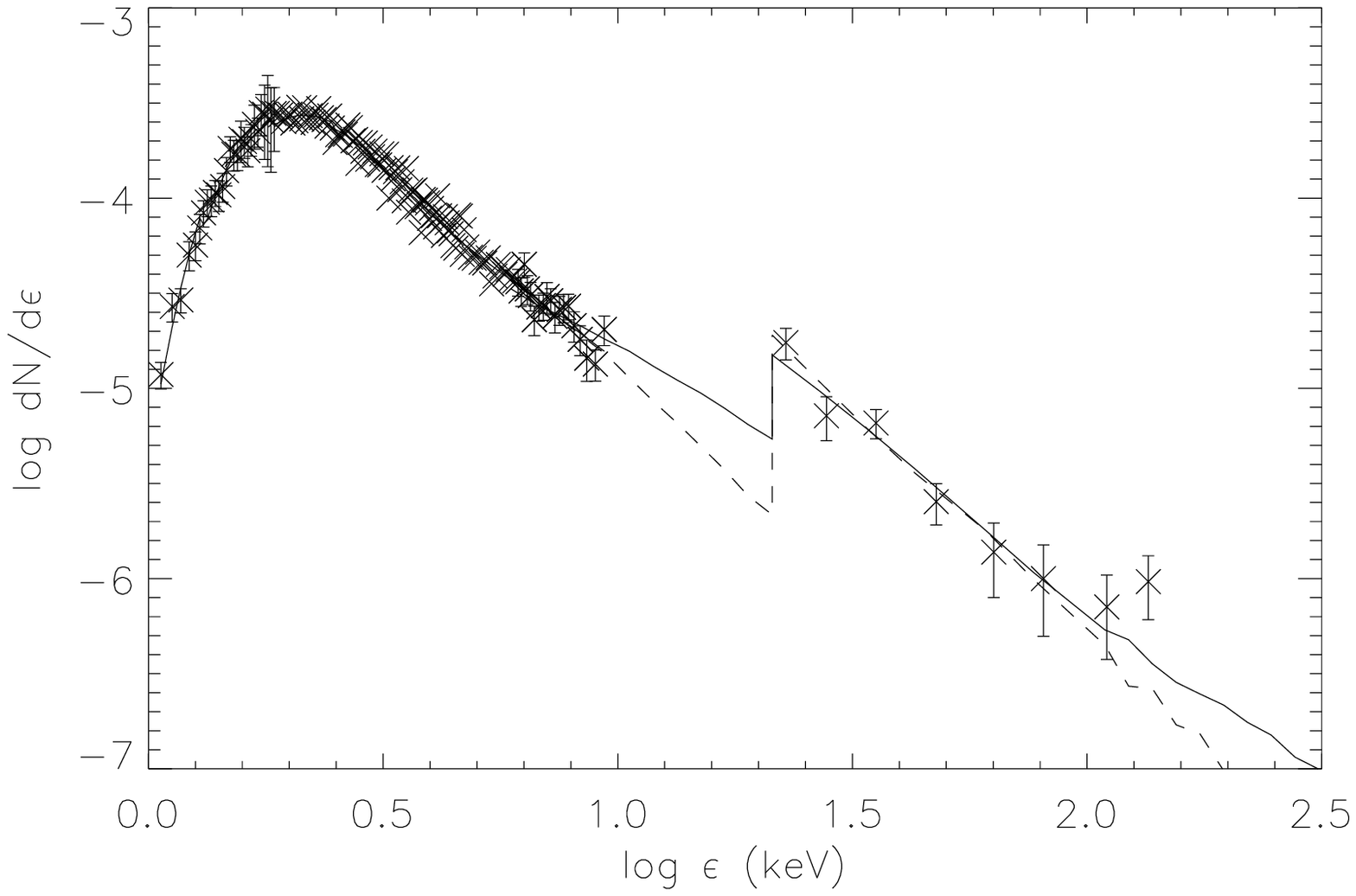,height=8.cm}}}
\caption{Combined \XMM\ and \INT\ spectrum of \sgrb. The dashed line is
the NTZ spectrum computed for the same values of parameters that best-fit
the \XMM\ data only (see Table
~\ref{tablexmmnonvar}), but using the QED cross sections in the Monte
Carlo code. This spectrum is than used to fit the  \INT\ data alone by 
leaving
free only the value of the normalization. This gives  $\chi^2=1.14$ for
7 d.o.f. and
the ratio between the \INT\ and the \XMM\ normalizations is 8.8. Solid
line: same as before but with  $\beta_{bulk} = 0.61$, $kT = 0.57 $~keV and
$\Delta \phi = 0.54 $; all parameters are at $3\sigma$ from their best
fitting values of the \XMM\ spectrum. Here the reduced  $\chi^2=1.07$
for 7 d.o.f. and
the ratio between the \INT\ and the \XMM\ normalizations is 2.8.}
\label{intfit} \end{figure*}

\end{document}